\documentclass[conference,compsoc]{IEEEtran}
\usepackage{hyperref}

\usepackage[justification=raggedright,format=hang]{caption}

\usepackage{url}
\makeatletter
\g@addto@macro{\UrlBreaks}{\UrlOrds}
\makeatother

\usepackage{cite}
\usepackage{amsmath,amssymb,amsfonts}
\usepackage{algorithmic}
\usepackage{graphicx}
\usepackage{textcomp}
\usepackage{float}
\usepackage{xcolor}
\def\BibTeX{{\rm B\kern-.05em{\sc i\kern-.025em b}\kern-.08em
    T\kern-.1667em\lower.7ex\hbox{E}\kern-.125emX}}
    
\usepackage{epigraph}
\usepackage{listings}
\usepackage{adjustbox}
\usepackage{multirow}
\usepackage{array}
\usepackage{authblk}
\usepackage[ruled,vlined]{algorithm2e}

\begin{document}

\title{FuzzSplore: Visualizing Feedback-Driven Fuzzing Techniques}

\author[1]{Andrea Fioraldi}
\author[1]{Luigi Paolo Pileggi}
\affil[1]{Sapienza University, Rome, Italy \authorcr {\tt \{fioraldi.1692419, pileggi.1691249\}@studenti.uniroma1.it}\vspace{1.5ex}}

\maketitle

\begin{figure*}[t]
\centering
\includegraphics[width=1.0\textwidth]{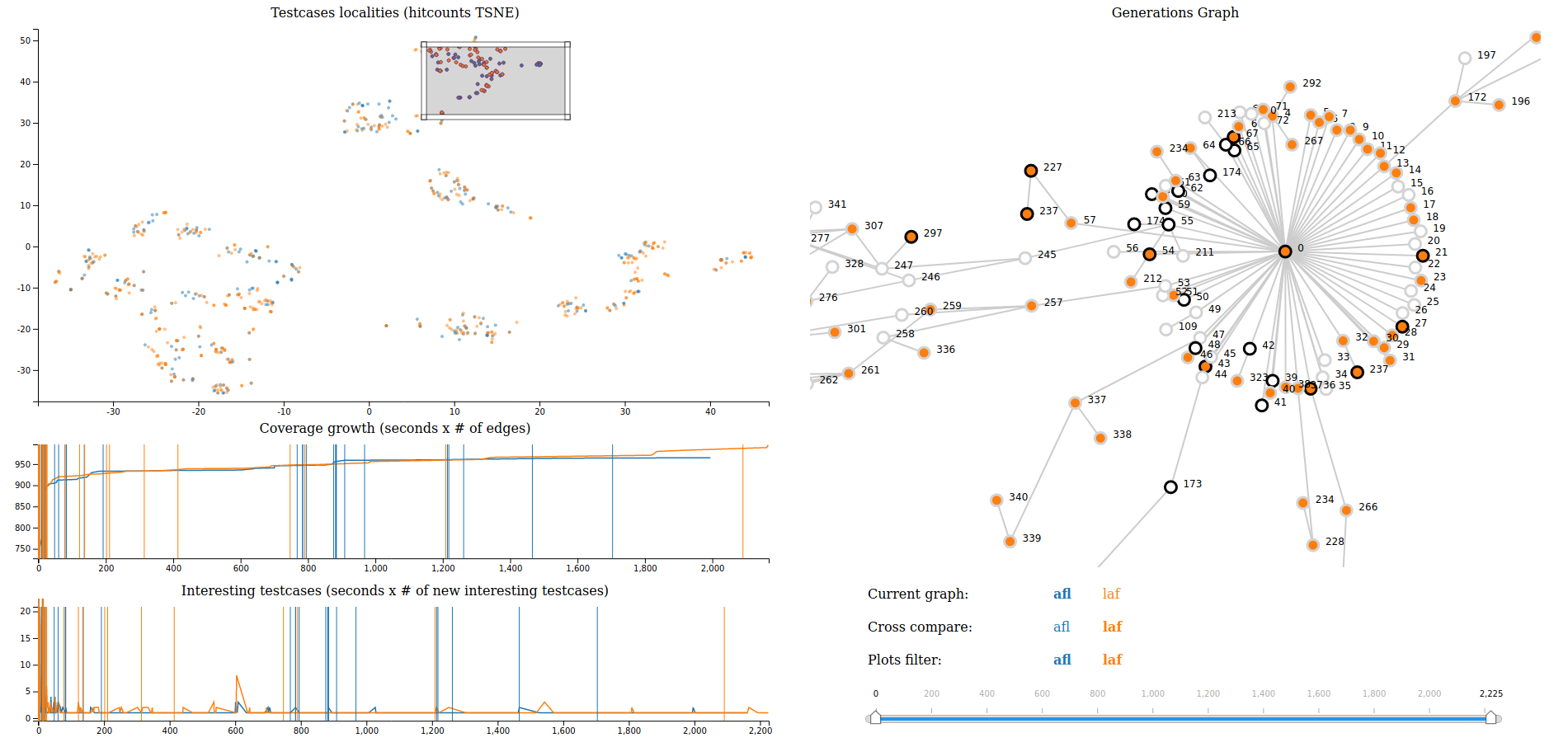}
\caption{Complete view of the {\sc FuzzSplore} visual panel.}
\label{fig:full}
\end{figure*}

\begin{abstract}
Fuzz Testing techniques are the state of the art in software testing for security issues nowadays.
Their great effectiveness attracted the attention of researchers and hackers and involved them in developing a lot of new techniques to improve Fuzz Testing.
The evaluation and the cross-comparison of these techniques is an almost open problem. In this paper, we propose a human-driven approach to this problem based on information visualization.
We developed a prototype upon the {\sc AFL++} fuzzing framework, {\sc FuzzSplore}, that an analyst can use to get useful insights about different fuzzing configurations applied to a specific target in order to choose or tune the best technique during a fuzzing campaign.
\end{abstract}

\section{Introduction}

{\em Fuzz Testing} or {\em Fuzzing} is a family of techniques to automatically uncovers bugs in software.

Due to its effectiveness, much more efficient than other software testing techniques like {\it Symbolic Execution} \cite{redqueen} \cite{sebastian}, the research in this field is flourishing and several different techniques were developed to improve fuzz testing, both from academia and industry.

The evaluation and the comparison of these techniques, however, is a debatable matter \cite{fuzzeval}.

A common proxy is the comparison of the code coverage reached over time by each fuzzer, due to the fact that a fuzzer cannot find a bug if it does not explore at least the vulnerable code segment.
Another widely used metric is found bugs over time, but a bug can be found just thanks to randomism or by specific target-dependant actions taken by the fuzzer and this makes the evaluations very prone to overfitting.

The data collected using these metrics are often representable using a simple time-based graph that shows the evolution of the fuzzing algorithm.

This approach is useful for immediate basic comparison between two or more techniques, an analyst has to just see which technique reaches more coverage in less time but does not reveal the properties of a fuzzer regards specific types of program states.

For instance, a technique can be better than another in exploring some types of program states and at the same time reaching less code coverage.
The technique will not cover the bugs in the unexplored code of course, but it may uncover bugs in the program points that it can better explore.
An example of such technique is the directed fuzzer towards sanitizers violations by \"Osterlund et al. \cite{parmesan}.

The problem of the evaluation of fuzzing techniques is important not only when the aim is to generally states which fuzzer is best, but also when an analyst wants to select the best fuzzers for a single target. It is common that fuzzers that are considered generally better than others on some targets perform worst than the others \cite{aflplusplus}.

We propose {\sc FuzzSplore}, a tool that allows an analyst to manually explore the evolution of different fuzzing techniques regards a single target program.

The main insight that a user can get using the tool are:

\begin{enumerate}
\item The ability of a fuzzer to generate clusters of inputs that are correlated in terms of covered program points;
\item The ability of a fuzzer in generating diversified inputs with its mutational algorithm;
\item The ability of a fuzzer to reach program points exploring intermediate inputs that are not an improvement in terms of coverage \cite{besensitive}.
\end{enumerate}

These insights can drive the user to choose the best technique to use for the selected program under test (PUT).

\section{Background}
\label{sec:back}

The simplest description of Feedback-driven Fuzzing is an algorithm that provides apparently random data to a computer program and then it watches for crashes or unexpected states and also saves the generated input for later processing if they cover interesting new states in terms of the chosen feedback \cite{fuzzing-book}.

Typically, the property of the program used as feedback is the set of the edges in the program Control Flow Graph \cite{compilerbook} in what is the so-called {\it Coverage-guided Fuzzing} (CGF).

\begin{figure}[H]
\includegraphics[scale=0.2]{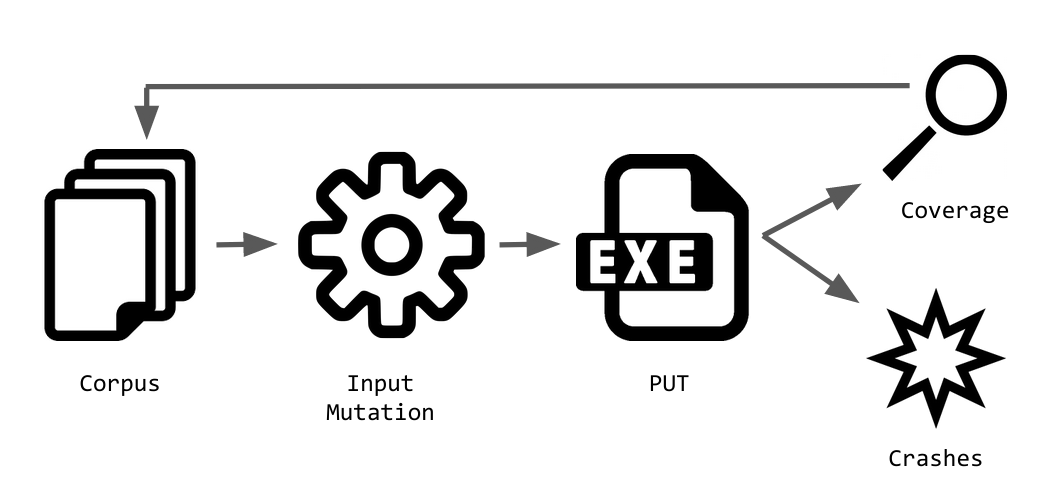}
\centering
\caption{Basic representation of the Mutational Coverage Guided Fuzzing algorithm}
\label{fig:cfg}
\end{figure}

The inputs are mutations of previously saved inputs in the fuzzing loop in Mutational Fuzzing (Figure \ref{fig:cfg}) or generated from scratch from a model in Generational Fuzzing.

We base our implementation on {\sc AFL++} \cite{aflplusplus}, a widely used fuzzer in recent times, that is a Mutational Coverage Guided Fuzzer.

State of the art Coverage-Guided Fuzzers encodes the approximate executed path in a representation that is easy and fast to process.
{\sc AFL++} uses a vector of 65536 entries by default, the {\it hitcounts} vector.

Each coordinate is associated with an edge and each value represents how many times the edge is executed modulo 256.

When a value greater than the previous one is registered in this vector, the fuzzer considers the input interesting and saves it.

Some extensions of CGF save also intermediate inputs that are a superset of the coverage reported in the hitcounts vector, like \cite{lafintel} \cite{ijon} \cite{besensitive}.

In general, when an input is saved, we can associate it to the testcases that generated it by mutation, the {\it parent} testcases. In this way, is easy to construct a graph of generated inputs that represents the progress of the hill-climbing algorithm of the fuzzer, the {\it Generations} Graph.


\section{Methodology}

A {\em fuzzing campaign} is the process of running one or more fuzzers for a long period of time or even continuously like in OSS-Fuzz \cite{serebryany2017oss}.

Security researchers typically start fuzzing using naive configurations and off-the-shelf fuzzers, then, meanwhile, the campaign runs, observe the evolution and tune the fuzzers.

Our proposed approach aims to insert in the observation-tuning feedback loop a visual component to help the researcher better understand insights about the fuzzers testing a particular target.

The data processed by {\em FuzzSplore} comes from the execution of the corpus of testcases that each fuzzer saved so far. The execution is instrumented and various properties are observed.

Then we visualize these collected properties and the user can relate them to better understand what is going. After that, the user can choose to drop some fuzzers if less effective and assign more resources (typically CPUs) to the most effective fuzzers or tune each individual fuzzer.

The fuzzing campaign can then continue. When it saturates, the analyst can collect insights using our tool and restart the visual analytics feedback-loop.

Saturation of fuzzers, when no more additional state is explored or the number of states explodes, is a problem that was rarely addressed in academic literature but that affects each type of Feedback-driven Fuzzer \cite{saturation},  and a tool that can guide towards the selection of techniques that avoid saturation can help a lot the campaign.

\subsection{Data Retrieval}


We denote each fuzzer $F_i$ where $i$ is the index that identifies it. 
With $PUT_i$ we denote the version of the PUT preprocessed and instrumented in order to be used by $F_i$.
$PUT_e$ is the version of the PUT that logs the edge coverage using the hitcounts vector. It has to be provided independently if it is used or not by some fuzzer $F_i$.
With $T_i(t)$ we denote the set of the saved testcases, the queue, by $F_i$ until time $t$ (seconds).

Given $t$ as the time chosen by the user to observe the progress of the fuzzers, the Algorithm \ref{alg:data} computes the following sets:

\begin{itemize}
\item the set $C$ of all the functions $C_i \colon Time \longrightarrow NumEdges$ that relates, for the fuzzer $F_i$, a time unit to the number of discovered edges so far;
\item the set $I$ of all the functions $I_i \colon Testcase \longrightarrow \{F_j, ...\}$ that associates, for the fuzzer $F_i$, each testcase in $T_i(t)$ to the set of fuzzers that consider the testcase as interesting;
\item the set $X$ of the sets $X_i$, that maintains, for each fuzzer, the hitcounts vectors associated with the execution of each testcase in $T_i(t)$;
\end{itemize}

\begin{algorithm}[h]
\DontPrintSemicolon

  \For {$F_i$ \textbf{ in } $Fuzzers$}{
    $V_{acc} \gets (0_0 ... 0_{65536}) $\;
    
    \For {$T$ \textbf{ in } $T_i(t)$}{
      $V \gets Execute(PUT_e, T)$\;
      $X_i \gets X_i \cup \{V\}$\;
      $V, IsInteresting \gets MergeCoverage(V_{acc}, V)$\;
      \If {$IsInteresting$}{
        $C_i(Time(T)) \gets CountNotZeros(V_i)$\;
      }
    }
    
    \For {$F_j$ \textbf{ in } $Fuzzers \setminus F_i$}{
    
     $V_{acc} \gets (0_0 ... 0_{65536}) $\;
     
      \For {$T$ \textbf{ in } $T_i(t)$}{
        $V \gets Execute(PUT_j, T)$\;
        $V_{acc}, IsInteresting \gets MergeCoverage(V_{acc}, V)$\;
        \If {$IsInteresting$}{
          $I_i(T) \gets I_i(T) \cup \{F_j\}$\;
        }
      }
    
    }
    
  }

 \Return{$C, I, X$}\;
 \caption{Compute $C$, $I$, and $X$}
 \label{alg:data}
\end{algorithm}

The next item that has to be retrieved, in addition to $C$, $I$ and $X$, is the set $G$ of all the graphs $G_i$ that describes the evolution of each $T_i(t)$, the levels graph introduced in Sec. \ref{sec:back}.

We assume that each fuzzer encodes the information about the parent testcases into the metadata of each testcase. In this way, it is trivial to construct the graph just by reading all the metadata in $T_i(t)$.

\subsection{Visualization}

We visualize the computed data $C$, $I$, $X$, $G$, and some other properties that can be directly collected in four different views.

You can see these views with some example data in the screenshot of our implementation, in Figure \ref{fig:full}.

A time bar is used to select $t' \in [0, t]$ to ignore data outside the selected time range and, for instance, visualize the data related to the queue $T_I(t')$ without the need to run again Algorithm \ref{alg:data}.

\subsubsection{Testcases Scatterplot}

Each $X_i$ is a matric of $|T_i(t)|$ rows in which each row is a vector of 65536 entries.

These raw numbers are raw to visualize. To handle this problem, we reduce the dimensionality of each vector $X_{i,j}$ from 65536 to 2, in order to be easily visualized in a scatterplot.

To do that, we chosen an algorithm that optimizes the conservation of local distances after the dimensionality reduction, {\em t-SNE} \cite{maaten2008visualizing}. The nature of this algorithm is random, it needs to process $X$ entirely in order to get new vectors that are meaningfully comparable.

We experimentally observed on a test dataset that a perplexity of 30 is good enough.

The user can select groups of nodes interactively to highlight properties in the other visualizations.

\subsubsection{Coverage Growth Plot}

$C$ can be visualized simply using a line plot with the X axis representing the domain, the time, and the Y axis the number of edges.

When a testcase is selected in the scatterplot or in the generations graph a vertical line appears at position $x$ where $x$ is the time in which the testcase was discovered.

\subsubsection{Interesting Testcases Plot}

This plot is used to visualize the evolution of the fuzzing algorithm in finding new testcases. The X axis represents time in seconds, the Y axis the number of new interesting testcases saved by the fuzzer in that second.
This information is directly contained in $T_i(t)$.

Here too, when a testcase is selected in the scatterplot or in the generations graph a vertical line appears at position $x$ where $x$ is the time in which the testcase was discovered.

\subsubsection{Generations Graph}

We visualize each Generations Graph $G_i$ combined with $I$. Given a fuzzer $F_j$ from the user, we highlight in graph $G_i$ each node associated with each testcase $T$ if $F_j \in I_i(T)$. In this way, the user can know if the evolution of $T_i(t)$ associated with the fuzzer $F_i$ is compatible with the selected $F_j$.

When a testcase is selected in the scatterplot, the border of the corresponding node in the graph is highlighted. The user can select additional nodes or deselect nodes selected from the scatterplot. The scatterplot selection is synchronized in both ways with the graph.

\subsection{Analyst Feedback}

The insights that an analyst can retrieve in order to choose or tune the fuzzers using the visualization are, but not limited to, the following:

\begin{itemize}
\item Looking just at the scatterplot, the user can select a subset of fuzzers that explore different program points if the points related to each fuzzer in the graphs are clustered;
\item Looking at the scatterplot and the coverage graph, the user can select a cluster of testcases that are similar and see the ability of a fuzzer in generating similar testcases in a small range of time. A fuzzer that discovers few points at a time and have them distributed for all the X axis of the coverage plot should be deprioritized;
\item Looking at the coverage graph, when there is a huge increment of the number of edges, the user can see if an outlier in the scatterplot was generated. This allows to isolate interesting testcases that improves a lot the coverage;
\item Selecting testcases in the graph, the user can see if the testcases are similar in the scatterplot in order to understand the ability of the mutator to generate similar or different derived inputs;
\item Selecting testcases in the graph and a fuzzer to cross-compare, the user can know if the coverage metric of the other fuzzer is sensitive enough to cover the selected testcases.
\end{itemize}

With this information, the security researcher should be able to choose and tune the set of fuzzers to avoid the saturation of the fuzzing campaign. This methodology is a first step towards a fuzzers debugger that is highly demanded by the security research community.


\section{Implementation}

We created an \textsc{html} page comprised of 4 views and a filtering panel and all the components were created using the \textsc{D3.js} library.

\subsection{Testcases Scatterplot}
 
\begin{figure}[h]
  \includegraphics[scale=0.25]{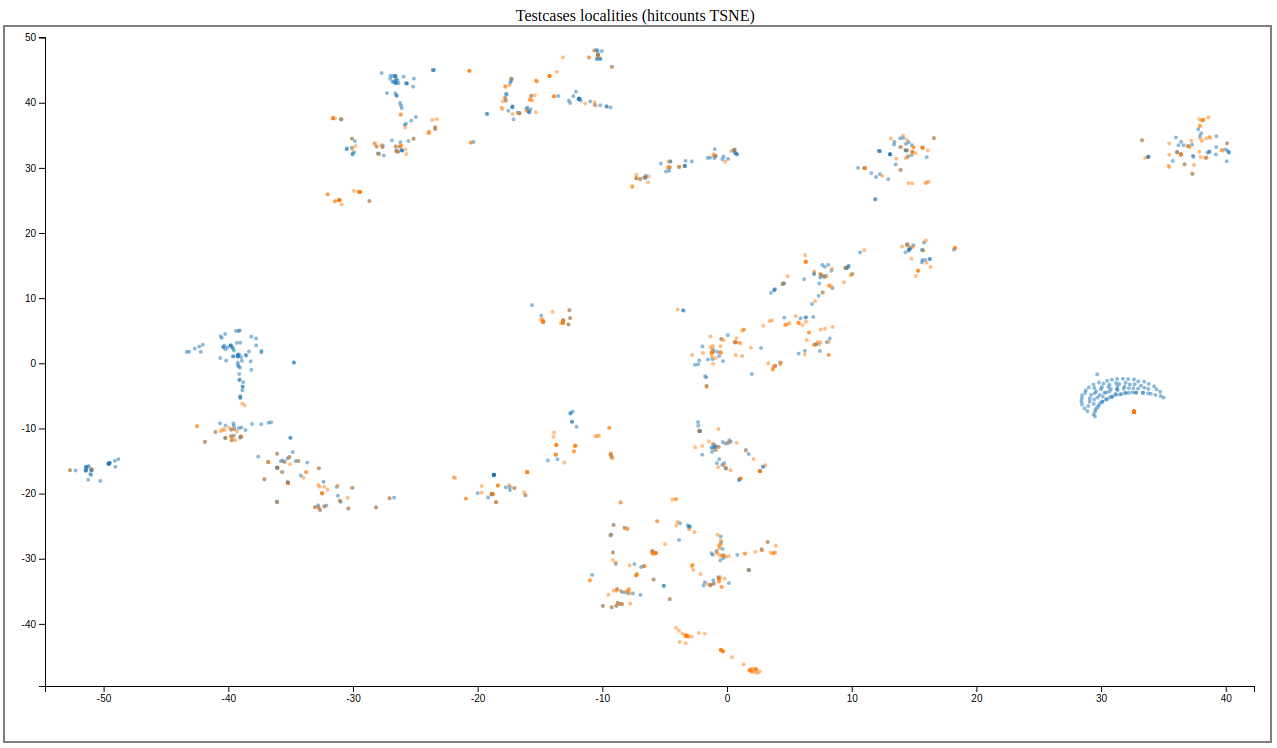}
  \centering
  \caption{Testcases scatterplot}
  \label{fig:scatterplot}
\end{figure}

The scatterplot (Fig. \ref{fig:scatterplot}) has, as both axes, a linear scale where the points are color-coded to represent a category to help the analyst distinguish the similarity in the clusters highlighted and the presence of outliers.\par
By brushing over the scatterplot a routine is called to update the other 3 views with the highlighted elements by selecting the corresponding nodes in the Generation graph and inserting lines in both plots.\par
The user can also zoom in and out and both axes are scaled appropriately.

\subsection{Coverage Growth and Interesting Testcases Plots}

\begin{figure}[h]
  \includegraphics[scale=0.25]{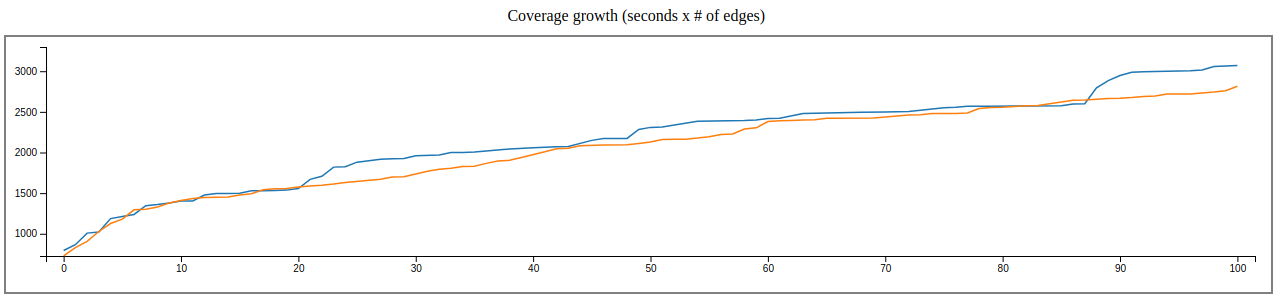}
  \centering
  \caption{Coverage growth plot}
  \label{fig:coverage}
  \includegraphics[scale=0.25]{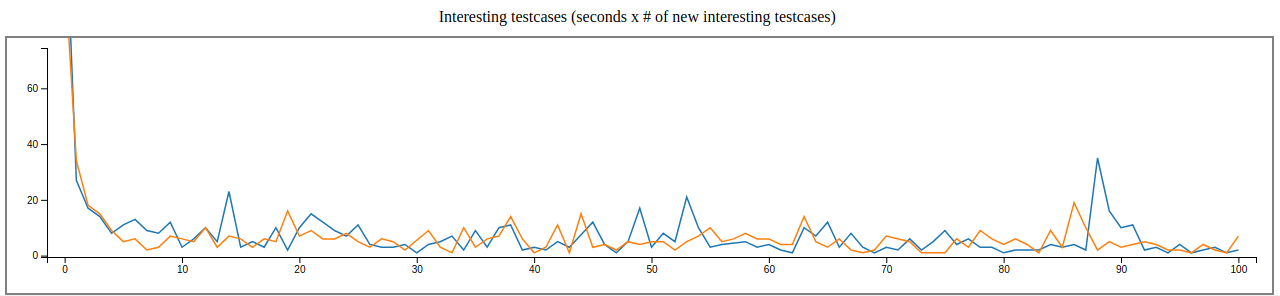}
  \centering
  \caption{Interesting Testcases plot}
  \label{fig:inputs}
\end{figure}

The Coverage graph (Fig. \ref{fig:coverage}) plots the growth over time of the number of covered edges, the Interesting Testcase graph (Fig. \ref{fig:inputs}) plots the number of new interesting test cases over time instead, for both the bottom axis is implemented as a linear scale, for the first graph the left axis is implemented as a logarithmic scale, for the latter a linear scale is used instead.\par
When data is selected on the scatterplot or graph vertical lines appear in both plots at the corresponding time having the stroke color matching the fuzzing technique.\par
We also implemented a pan and zoom functionality that keeps the lowest value pinned at the bottom.

\subsection{Generation Graph}
\begin{figure}[H]
  \includegraphics[scale=0.25]{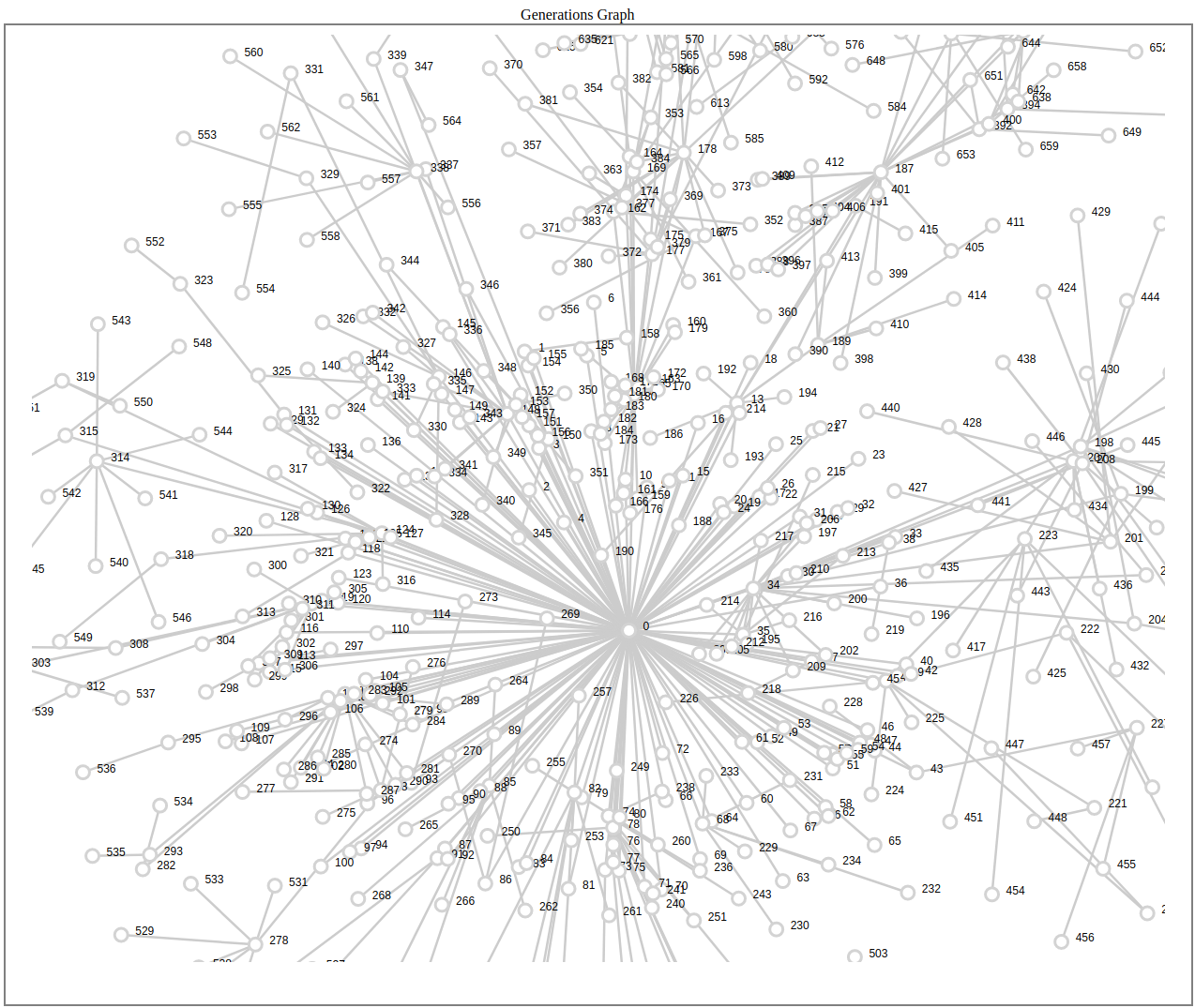}
  \centering
  \caption{Generation graph}
  \label{fig:tree}
\end{figure}

The Generation Graph (Fig. \ref{fig:tree})  is created as a hierarchical layout where each data point's value is displayed as a node label.\par
The user can zoom as well as pan over the entire view to have a better understanding of the data and when a node is selected in the other a routine is called to highlights the corresponding points in the scatterplot and insert lines in the other plots.
A mouseover on node lowlights all the nodes except the hovered nodes and their neighbor nodes and edges.

\subsection{Filtering Panel}
\begin{figure}[h]
  \includegraphics[scale=0.25]{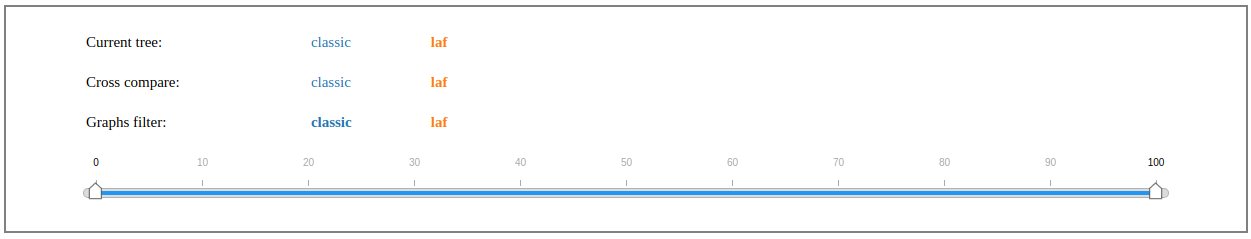}
  \centering
  \caption{Filtering panel}
  \label{fig:panel}
\end{figure}

The user can filter the data shown an all the 4 views by time, with a range slider (Fig. \ref{fig:panel}) located at the bottom right of the page, and by category, by clicking on the category names directly on top of the slider, the filtering works by updating the existing graphs without redrawing.

\section{Concluding Remarks}

{\sc FuzzSplore} brings a useful visualization-based method to retrieve insights from running fuzzers in a campaign. 

It defines a long-term visual analytics feedback loop applied to fuzzing with a set of data retrieval and visualization techniques that can be easily extended in future works. 

The information that a security researcher can collect using our approach can help in understanding the problem of saturation in fuzzing campaigns, a widely spread problem that is rarely addressed in academic literature.

We share {\sc FuzzSplore} as Free and Open Source Software at \url{https://github.com/andreafioraldi/FuzzSplore}.

\vspace{10mm}

\bibliographystyle{IEEEtran}
\bibliography{bibliography}

\begin{thebibliography}{10}
\providecommand{\url}[1]{#1}
\csname url@samestyle\endcsname
\providecommand{\newblock}{\relax}
\providecommand{\bibinfo}[2]{#2}
\providecommand{\BIBentrySTDinterwordspacing}{\spaceskip=0pt\relax}
\providecommand{\BIBentryALTinterwordstretchfactor}{4}
\providecommand{\BIBentryALTinterwordspacing}{\spaceskip=\fontdimen2\font plus
\BIBentryALTinterwordstretchfactor\fontdimen3\font minus
  \fontdimen4\font\relax}
\providecommand{\BIBforeignlanguage}[2]{{%
\expandafter\ifx\csname l@#1\endcsname\relax
\typeout{** WARNING: IEEEtran.bst: No hyphenation pattern has been}%
\typeout{** loaded for the language `#1'. Using the pattern for}%
\typeout{** the default language instead.}%
\else
\language=\csname l@#1\endcsname
\fi
#2}}
\providecommand{\BIBdecl}{\relax}
\BIBdecl

\bibitem{redqueen}
\BIBentryALTinterwordspacing
C.~Aschermann, S.~Schumilo, T.~Blazytko, R.~Gawlik, and T.~Holz, ``{REDQUEEN:}
  fuzzing with input-to-state correspondence,'' in \emph{26th Annual Network
  and Distributed System Security Symposium, {NDSS}}, 2019. [Online].
  Available:
  \url{https://www.ndss-symposium.org/ndss-paper/redqueen-fuzzing-with-input-to-state-correspondence/}
\BIBentrySTDinterwordspacing

\bibitem{sebastian}
\BIBentryALTinterwordspacing
S.~Poeplau and A.~Francillon, ``Systematic comparison of symbolic execution
  systems: Intermediate representation and its generation,'' in
  \emph{Proceedings of the 35th Annual Computer Security Applications
  Conference}, ser. ACSAC ’19.\hskip 1em plus 0.5em minus 0.4em\relax New
  York, NY, USA: Association for Computing Machinery, 2019, p. 163–176.
  [Online]. Available: \url{https://doi.org/10.1145/3359789.3359796}
\BIBentrySTDinterwordspacing

\bibitem{fuzzeval}
\BIBentryALTinterwordspacing
G.~Klees, A.~Ruef, B.~Cooper, S.~Wei, and M.~Hicks, ``Evaluating fuzz
  testing,'' in \emph{Proceedings of the 2018 ACM SIGSAC Conference on Computer
  and Communications Security}, ser. CCS ’18.\hskip 1em plus 0.5em minus
  0.4em\relax New York, NY, USA: Association for Computing Machinery, 2018, p.
  2123–2138. [Online]. Available:
  \url{https://doi.org/10.1145/3243734.3243804}
\BIBentrySTDinterwordspacing

\bibitem{parmesan}
\BIBentryALTinterwordspacing
S.~Österlund, K.~Razavi, H.~Bos, and C.~Giuffrida, ``{ParmeSan}:
  {Sanitizer}-guided {Greybox} {Fuzzing},'' in \emph{{USENIX} {Security}}, Aug.
  2020. [Online]. Available:
  \url{Paper=https://download.vusec.net/papers/parmesan_sec20.pdf
  Code=https://github.com/vusec/parmesan}
\BIBentrySTDinterwordspacing

\bibitem{aflplusplus}
\BIBentryALTinterwordspacing
A.~Fioraldi, D.~Maier, H.~Ei{\ss}feldt, and M.~Heuse, ``Afl++ : Combining
  incremental steps of fuzzing research,'' in \emph{14th {USENIX} Workshop on
  Offensive Technologies ({WOOT} 20)}.\hskip 1em plus 0.5em minus 0.4em\relax
  {USENIX} Association, Aug. 2020. [Online]. Available:
  \url{https://www.usenix.org/conference/woot20/presentation/fioraldi}
\BIBentrySTDinterwordspacing

\bibitem{besensitive}
\BIBentryALTinterwordspacing
J.~Wang, Y.~Duan, W.~Song, H.~Yin, and C.~Song, ``Be sensitive and
  collaborative: Analyzing impact of coverage metrics in greybox fuzzing,'' in
  \emph{22nd International Symposium on Research in Attacks, Intrusions and
  Defenses ({RAID} 2019)}.\hskip 1em plus 0.5em minus 0.4em\relax Chaoyang
  District, Beijing: {USENIX} Association, Sep. 2019, pp. 1--15. [Online].
  Available: \url{https://www.usenix.org/conference/raid2019/presentation/wang}
\BIBentrySTDinterwordspacing

\bibitem{fuzzing-book}
A.~Zeller, R.~Gopinath, M.~B\"{o}hme, G.~Fraser, and C.~Holler, ``{The Fuzzing
  Book},'' \url{https://www.fuzzingbook.org/}, 2019, [Online; accessed
  10-Sep-2019].

\bibitem{compilerbook}
K.~Cooper and L.~Torczon, \emph{Engineering a Compiler: International Student
  Edition}.\hskip 1em plus 0.5em minus 0.4em\relax San Francisco, CA, USA:
  Morgan Kaufmann Publishers Inc., 2003.

\bibitem{lafintel}
``{Circumventing Fuzzing Roadblocks with Compiler Transformations},''
  \url{https://lafintel.wordpress.com/2016/08/15/circumventing-fuzzing-roadblocks-with-compiler-transformations/},
  2016.

\bibitem{ijon}
C.~Aschermann, S.~Schumilo, A.~Abbasi, and T.~Holz, ``Ijon: Exploring deep
  state spaces via fuzzing,'' in \emph{IEEE Symposium on Security and Privacy
  (Oakland)}, 2020.

\bibitem{serebryany2017oss}
K.~Serebryany, ``Oss-fuzz-google’s continuous fuzzing service for open source
  software,'' in \emph{USENIX Security Symposium}, 2017.

\bibitem{saturation}
A.~Groce and J.~Regehr, ``{The Saturation Effect in Fuzzing},''
  \url{https://blog.regehr.org/archives/1796}.

\bibitem{maaten2008visualizing}
L.~v.~d. Maaten and G.~Hinton, ``Visualizing data using t-sne,'' \emph{Journal
  of machine learning research}, vol.~9, no. Nov, pp. 2579--2605, 2008.

\end{thebibliography}

\end{document}